\documentclass[12pt]{article}

\usepackage{amssymb,amsmath,amsfonts,eurosym,geometry,ulem,graphicx,caption,color,setspace,sectsty,comment,footmisc,caption,pdflscape,subfigure,array,hyperref}

\usepackage[square,sort,comma,numbers]{natbib}

\newcolumntype{L}[1]{>{\raggedright\let\newline\\arraybackslash\hspace{0pt}}m{#1}}
\newcolumntype{C}[1]{>{\centering\let\newline\\arraybackslash\hspace{0pt}}m{#1}}
\newcolumntype{R}[1]{>{\raggedleft\let\newline\\arraybackslash\hspace{0pt}}m{#1}}

\begin{document}

\begin{titlepage}
\title{The Doctrine of Cyber Effect: An Ethics Framework for Defensive Cyber Deception}
\author{Quanyan Zhu\thanks{Q. Zhu is with the Department of Electrical and Computer Engineering, New York University, 370 Jay Street, Brooklyn, NY; E-mail: qz494@nyu.edu}}
\date{\today}
\maketitle
\begin{abstract}
\noindent The lack of established rules and regulations in cyberspace is attributed to the absence of agreed-upon ethical principles, making it difficult to establish accountability, regulations, and laws. Addressing this challenge requires examining cyberspace from fundamental philosophical principles. This work focuses on the ethics of using defensive deception in cyberspace, proposing a doctrine of cyber effect that incorporates five ethical principles: goodwill, deontology, no-harm, transparency, and fairness. To guide the design of defensive cyber deception, we develop a reasoning framework, the game of ethical duplicity, which is consistent with the doctrine. While originally intended for cyber deception, this doctrine has broader applicability, including for ethical issues such as AI accountability and controversies related to YouTube recommendations. By establishing ethical principles, we can promote greater accountability, regulation, and protection in the digital realm.
%Cyberspace lacks established rules and regulations due to the absence of agreed-upon ethical principles. This makes it challenging to establish accountability, regulations, and laws. To address this, we must examine cyberspace from fundamental philosophical principles. This work focuses on the ethics of using defensive deception in cyberspace. We propose a doctrine of cyber effect, which incorporates five ethical principles: goodwill, deontology, no-harm, transparency, and fairness. We also develop a reasoning framework called the game of ethical duplicity, which is consistent with the doctrine and can help guide the design of defensive cyber deception. Although our doctrine was originally intended for use in cyber deception, it can also be applied to other ethical issues such as AI accountability and controversies surrounding YouTube recommendations. By establishing ethical principles, we can promote greater accountability, regulation, and protection in the digital realm. \\
\vspace{0in}\\
%\noindent\textbf{Keywords:} key1, key2, key3\\
%\vspace{0in}\\
%\noindent\textbf{JEL Codes:} key1, key2, key3\\

\bigskip
\end{abstract}
\setcounter{page}{0}
\thispagestyle{empty}
\end{titlepage}
\pagebreak \newpage

\doublespacing

\section{Introduction and Motivation} \label{sec:introduction}

Defensive cyber deception design is a strategy that aims to develop and implement deception technologies, such as honeypots \cite{huang2019adaptive,spitzner2003honeypots} and moving target defense \cite{zhu2013game,jajodia2011moving}, to protect computing systems from increasingly sophisticated attackers. The ultimate objective of this design is to optimize security performance metrics, including the probability of entrapment of attackers and the success rate of strategic defense measures. The use of defensive cyber deception design has been successful in a variety of contexts, from corporate networks to government agencies. Organizations have implemented a number of deception-based security measures to protect its networks and systems.  

One area of concern is the use of deception against employees or internal stakeholders. For instance, organizations may utilize deceptive tactics to detect insider threats or assess the security awareness of employees. While these tactics can be effective, they may also be seen as a breach of trust and privacy, potentially damaging the relationship between the organization and its employees. Even when deception is employed against attackers or illegitimate users, ethical concerns remain. This is particularly relevant in the case of honeypots, where employees may be lured into the trap and falsely classified as insiders upon entrapment. The use of misleading signals or employee curiosity can also cause non-insiders to fall into honeypot traps. Such practices can further erode trust and employee morale, raising ethical questions about the use of deception in cybersecurity.

The other ethical concern with cyber deception is the use of deception against attackers. While it is generally accepted that organizations have the right to defend themselves against cyber attacks, the use of deception techniques such as honeypots may be seen as unfair or unethical. It can be argued that it manipulates the attacker into taking an action they may not have taken otherwise. Essentially, honeypots are computer systems designed to look like legitimate targets but are actually decoys set up to detect or deflect cyberattacks. By intentionally making the intruders believe that they have successfully compromised a system, honeypots can mislead the intruders into taking actions that they would not take otherwise.

\begin{figure}[h]
\centering
 \includegraphics[width=.6\linewidth]{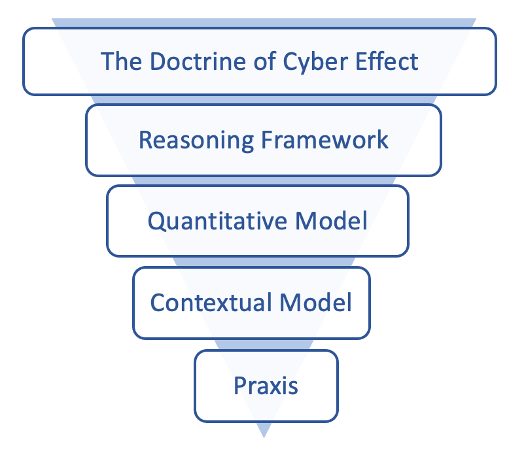}
 \caption{The top-down application of the Doctrine of Cyber Effect to cybersecurity problems: The Games of Ethical Duplicity (GED) offer a framework for reasoning that is in line with ethical doctrine. This framework can be formalized using mathematical models for both quantitative analysis and design. The resulting quantitative model is customizable and can be applied to specific contexts within network security, such as insider threats and advanced persistent threats. This enables the creation of contextual models, which in turn make it possible to implement computational methods, protocols, and data analytics for practical use.}
\label{fig:ethics}
\end{figure}

An ethics framework is necessary to provide guiding principles for defining ethical norms and creating design methodologies that align with these principles. This need is particularly relevant in the field of defensive cyber deception (DCD) \cite{pawlick2019game,huang2019dynamic}. In this work, we discuss the recent developments in defensive cyber deception and propose the doctrine of cyber effect as a means of stipulating principles that are crucial to defining ethical norms. The doctrine of cyber effect provides a set of principles that can guide the ethical use of deception in cyberspace. Additionally, this work explores game-theoretic design methodologies that align with the doctrine of cyber effect, and their applications in mitigating insider threats. By providing an ethical framework and design methodologies that are consistent with it, we can promote the responsible use of cyber deception techniques while minimizing potential ethical violations. This discussion is conceptualized using Figure \ref{fig:ethics}.

This paper follows the following organizational structure to discuss the ethical implications of defensive cyber deception. Specifically, Section \ref{sec:DCD} introduces the concept of defensive cyber deception and highlights its unique characteristics. In Section \ref{sec:ethics}, we review classical ethics frameworks to provide a theoretical basis for examining the ethical considerations of defensive cyber deception. Section \ref{sec:doctrine} then elaborates on the doctrine of cyber effects and its relevance to defensive cyber deception. Next, Section \ref{sec:reasoning} proposes a novel approach to developing a consistent ethical reasoning framework for defensive cyber deception through the use of games of ethical duplicity. Furthermore, in Section \ref{sec:broader}, we discuss the broader implications of the doctrine of cyber effects beyond the realm of defensive cyber deception. Finally, Section \ref{sec:conclusion} concludes the paper by summarizing the main points and highlighting the significance of ethical considerations in the practice of defensive cyber deception.

%This paper is organized as follows. Section \ref{sec:DCD} introduces defensive cyber deception and discusses its distinct features. Section \ref{sec:ethics} discusses classical ethics frameworks. The doctrine of cyber effects is elaborated in Section \ref{sec:doctrine}. Section \ref{sec:reasoning} proposes games of ethical duplicity to provide a consistent reasoning framework. Section \ref{sec:broader} discusses the relevance of the doctrine beyond the realm of defensive cyber deception. The paper is concluded in Section \ref{sec:conclusion}.

\section{Defensive Cyber Deception}\label{sec:DCD}

The recent advances in information technologies and their integration with the physical world have revolutionized autonomous driving, industry 4.0, VR/AR technologies, and beyond. This convergence has also brought new challenges to protect the system from being tampered with or unauthorized access. Traditional approaches to cyber defense increasingly are inadequate to defend against increasingly sophisticated attacks that attempt to persistently analyze, probe, circumvent or fool such mechanisms. These attacks are often backed with information and resource advantages. Lessons from recent Target data breach and SolarWinds attacks have indicated that cyber defenses in enterprise and government networks such as firewalls, malware scanners, and introduction and detection technologies can be evaded by advanced persistent threats (APTs) \cite{chen2017security,rass2016gadapt}. Hence it is important to tilt the information and resource asymmetries between the attacker and the defender. In addition, it is important to assume that an adversary will breach perimeter defenses and establish footholds within the defender's network. Defense mechanisms that can engage the adversary and influence the adversary's moves to the defender's advantage will strengthen the security and resilience of the network \cite{zhu2015game,huang2021reinforcement}. To this end, DCD has emerged as an effective paradigm of proactive security defense that deliberately confound the adversaries, increases the cost of their information gathering, and misleads them to take actions that are in favor of the defender.

Recent years have witnessed growing attention and interest in this space. Moving target defense (MTD) strategies have been developed to change, shift, or reduce the attack surface available, increasing the uncertainties and the costs for an attacker. Honeypot-based techniques are another class of mechanisms that are designed to provide deceptive information to attackers and proactively collect information about their means and objectives. Cyber deception has shown great promise to proactively disrupt the cyber kill chain and has been increasingly gaining interest within both academia and industry.  
The use of DCD can sometimes inadvertently attract innocent legitimate users who mistake them for legitimate systems. This can cause collateral damage if they mistake the honeypots for real systems and engage with them. They may be denied access to the services or resources that are being simulated by the honeypot. This can cause frustration and disrupt business operations. In some cases, they may inadvertently provide sensitive information to the honeypot, thinking that they are interacting with a legitimate system. This can result in the loss of confidential information, which can have serious consequences. Furthermore, if legitimate users discover that they have been interacting with a honeypot, they may lose trust in the organization that deployed it. 
The techniques of DCD can potentially be misused by organizations that may deploy them for offensive purposes, such as using the information collected from the honeypot to launch attacks on other systems or to steal sensitive information.  The data collected by the honeypot can be misused by an insider or an attacker to create more effective phishing attacks \cite{gupta2017fighting,van2019cognitive}, social engineering attacks \cite{salahdine2019social}, or other types of attacks that target specific individuals or organizations.

\subsection{Overt Deception vs. Covert Deception}

Information is the key component of the DCD. Its goal is to create an information advantage for the defender. There are two fundamental types of deception mechanisms characterized by the two polarizing information structures. One is the over deception, and the other one is the cover deception.  Overt deception refers to the scenario where the uncertainties and the policies are announced to the users. The announced policies are committed by the designer. The designer informs the users of the likelihood of honeypots but does not disclose which ones are honeypots.  Covert deception refers to the scenario where the designer can directly intercept and alter the legitimate message or send a deceptive message to a user. The user is not aware of the presence of the deception and acts as if he were in a non-adversarially prescribed setting. In the case of honeypot deployment, its operation is hidden, and network users access information and services without knowing the presence and likelihood of honeypots.

\subsection{Incentives and Objectives}

Another important component of cyber deception is the incentive or intent behind it. One common intent of using deception techniques is to detect and prevent attacks before they cause harm. Defenders can create decoys or honeypots to divert attackers away from their real targets and into a monitored environment where they can be studied and analyzed. However, it is important to note that an unethical intent of using cyber deception is to deliberately harm or deceive others. For instance, attackers may use decoys or honeypots with the goal of trapping legitimate or illegitimate users and infecting them with malware or stealing their sensitive data. Such actions are malicious and illegal. Another unethical use of deception techniques is when an organization uses them as an excuse to entrap innocent individuals for the purpose of obtaining their personal information for marketing purposes. This is a clear violation of ethical and legal norms as it breaches users' privacy and trust. It is crucial to use deception techniques ethically, with the intent of protecting systems and networks from potential attacks, while respecting the rights and privacy of users.

\subsection{Distinct features of DCD}

Cyber deception is distinct from other forms of deception, such as false advertising and impersonation in the physical world, because it involves creating a false reality in the digital domain. The cost of cyber deception is often lower, and it is more convenient to create and manage. However, its impact can be more immediate and widespread, as it can be used to disrupt or manipulate critical digital systems and networks. Despite the fact that DCD is designed to target attackers, it can unintentionally impact innocent users. For instance, in the context of honeypots, every user can potentially encounter one without knowing its exact location, and an innocent user can mistakenly access it. Therefore, leveraging Rawls' ``veil of ignorance" \cite{vermeule2001veil} is critical for an ethical design of DCD. It is important to consider the possibility that an innocent user can be trapped into a honeypot at the early stages of design (ex ante phase) and create a DCD mechanism that can inherently deal with consequence of the trapping of innocent users.

Deception strategies are not homogenous in nature; rather, their costs and potential for spillover effects can vary considerably. For instance, the deployment and configuration of honeypots may be more resource-intensive compared to other techniques, such as changing device addresses through the use of moving target defense. However, honeypots may lead to temporary denial of service attacks on legitimate users, or even the entrapment of innocent parties. Ultimately, the selection of a particular deception strategy must take into account the goals of the defender, as well as their awareness of the potential costs and effects of the strategy. In this sense, morality is an epistemic choice, as selecting the appropriate constraints can guide design choices towards a desirable ethical standard. These constraints can be either self-imposed, where ethical norms stem from the defender's moral principles, or externally imposed through compliance rules, regulations, or laws that reflect societal norms of morality. Regardless of the source of these constraints, their selection can shape the overall ethical landscape of the deception strategy, and ultimately impact the decision-making process.

Given the budgetary constraints of deploying a Deception-based Cyber Defense (DCD) solution, it may not always be possible to use the most effective deception technique to achieve ideal outcomes. Instead, there is a necessary trade-off between the outcomes that can be achieved and the moral constraints that must be considered. The imposition of more moral constraints may limit the potential of the DCD. Thus, the concept of ``moral budgets" has emerged, which represents the set of internal or external moral constraints that can be imposed, such as deontological rules or compliance regulations. The careful management and allocation of these moral budgets is crucial to ensuring a balance between the efficacy of the DCD and the ethical and legal considerations that must be respected.

\section{Classical Ethics Frameworks for Deception}\label{sec:ethics}

Kantian morality holds that the supreme principle of morality is a standard of rationality known as the ``Categorical Imperative" (CI). This objective, rationally necessary, and unconditional principle must always be followed, despite any natural desires or inclinations to the contrary. According to Kant \cite{kant2001lectures}, lying is never permissible, even if the intention is to bring about good consequences, such as lying to protect a victim from an attacker. We ought to avoid known wrongs, such as lying to avoid potential harm, because we cannot fully anticipate the consequences of our actions, and the result may be unexpectedly harmful. Nonetheless, we are blameless when we act according to our duty, even if there are harmful consequences. The focus of our categorical obligations is to keep our own agency free of moral taint, rather than on how our actions might cause or enable others to do evil. Kantian deontology creates a set of choices that are either morally required, forbidden, or permitted, with the rightness of a choice determined by its conformity with a moral norm. A Kantian agent guided by categorical imperatives is referred to as {\it homo moralis}.

The Kantian approach to ethics and morality prohibits the use of cyber deception in defense strategies, as deception involves providing misleading information to individuals who are not aware of the truth. Therefore, a Kantian defender would reject the idea of using cyber deception and instead rely on traditional passive approaches such as intrusion detection, cryptography, and training. However, in the modern cybersecurity landscape, these passive approaches have proven to be increasingly ineffective against sophisticated attackers. The elimination of cyber deception as an option for defense may not be practical or sound in light of this reality.

While the use of cyber deception may conflict with Kant's moral framework, it is worth considering whether the harm caused by cyber attacks justifies the use of deception as a means of preventing such attacks. This raises questions about the trade-offs between moral principles and practical considerations, and the extent to which defenders are justified in using potentially unethical means to achieve a desired end.

Kantian deontological ethical theory is contrasted with consequentialism \cite{de2017utilitarianism}, the doctrine that the morality of an action is to be judged solely by its consequences.  Utilitarianism, a consequentialist approach, holds that an action is morally right if it maximizes overall happiness or well-being, without regard to individual rights or justice. The carrier of this view, as mentioned, is often referred to as {\it homo economicus}.

While the Kantian principle clearly forbids deception, consequentialism may permit deception if it produces a positive outcome. Consequentialism argues that a homo economicus should plan for the best outcome no matter whether there is a need for hiding or manipulating information to all users as long as the network is secured. It does not restrain the arguably unethical behaviors of using honeypots, including the misuse of the data, creation of misleading information for innocent users, and the unconsented deployment of deception. Furthermore, how to define and evaluates benign consequences for DCD is a challenging within the consequentialism framework.
In the context of DCD, the limitations of both Kantian deontological ethics and consequentialism are evident. The Kantian approach prohibits the use of deception, which may limit the effectiveness of DCD in modern defense. On the other hand, consequentialism lacks actionable ethical principles for DCD and fails to provide a framework for evaluating the morality of such actions. Therefore, an ethical framework for DCD must take into account both the inherent value of moral actions and the potential consequences of such actions. It should aim to balance the benefits of DCD with the ethical concerns raised by the use of deception. Additionally, it should consider the rights and interests of all users, including innocent users who may be affected by the deployment of honeypots or other forms of deception. 

To create a new ethics framework, we can consolidate Kantian deontology into the framework for DCD by placing constraints on the actions and preferences of the agent, while the remaining freedom of choices is guided by utilitarian consequentialism. This approach creates a hybrid agent that combines the ethical principles of both deontology and consequentialism. Such an agent is essential in the DCD, where the designer has the power to create mechanisms that can incentivize and guide users or participants towards ethical behavior. Furthermore, the ethics framework for DCD should also support the principles of transparency, fairness, and the no-harm principle. Transparency means that the rules and mechanisms of the DCD should be known to all users, and legitimate or illegitimate users consent to the presence of DCD. Fairness ensures that the defense mechanism is designed to consider the needs and interests of all types of users, and not just some. The designer needs to be agnostic of the type of the users to which he belongs, and create policies that can accommodate all for the social good.  The no-harm principle has been advocated in many areas including medicine \cite{sharpe1997no}, environment \cite{mayer2016relevance}, and technology \cite{cerf2011first,winfield2019ethical}. In the context of DCD, it requires that the defender should not cause any harm to the legitimate users, and the design of the defense mechanism should minimize the risk of harm to anyone involved.

Social contract theory \cite{castiglione1995social} is a widely recognized ethical framework that aims to justify the existence of moral and political systems by proposing a hypothetical ``social contract" between individuals in a society. This theory suggests that individuals willingly relinquish some of their individual liberties and submit to a governing authority in return for protection of their basic rights and the advantages of living in a stable and well-ordered society. In the realm of ethics, social contract theory can be utilized to develop a set of moral principles that are rooted in the social contract, which can provide a basis for ethical decision-making.

Applying social contract theory to the context of DCD, we can see that users can voluntarily surrender some of their individual rights and choose to use the network with DCD in exchange for cyber protection and the advantages of living in a secure cyberspace. Furthermore, an attacker who chooses to attack the network with DCD does so with full knowledge that they could be caught and subject to scrutiny. The defender's ``agreement" with both the attacker and the users creates a set of mutual obligations and responsibilities. Users have a responsibility to abide by the rules established by DCD, while the defender has an obligation to enforce the rules. The attacker must explore the vulnerabilities of the network created by the defender, but must also be aware of the consequences of their actions.

The doctrine of cyber effects is rooted in the social contract theory, which emphasizes the importance of transparency in informing users, including attackers, of the policies related to the DCD network. This principle promotes the voluntary use and access of the network, with users being aware of the ``social contract" that comes with being part of the network. The agreement between the users and the network can be mathematically represented by binding constraints set by the agents. It is important to note that users may choose not to use the DCD network or find an alternative if they are not comfortable with the possibility of being mistakenly caught in a honeypot and having their legitimate access revoked. Similarly, attackers may choose to give up attacking the network if they become aware of the possibility of honeypots. The participation in the contract is voluntary for all agents involved. Thus, the doctrine of cyber effects creates an ethical standard based on the principles of the social contract theory. 

\section{The Doctrine of Cyber Effect: Five Principles for Ethical DCD}\label{sec:doctrine}

The classical ethical approaches, including those of Kant, utilitarianism, and virtue ethics, have limitations when it comes to designing Deception for Cyber Defense (DCD). Kant's approach would prohibit the design of DCD, as it considers deception to be morally impermissible regardless of its intended use. Meanwhile, utilitarianism lacks clear constraints and principles to guide the ethical use of DCD. The virtue approach, on the other hand, emphasizes character traits and intentions over actions. However, applying virtue ethics to the design of DCD can be challenging, as it is difficult to determine the right character traits and intentions that should guide the use of deception in cyberspace.

To address these challenges, we propose the doctrine of cyber effects, a framework to support the ethical design of DCD. This framework considers various ethical approaches and principles, such as moral constraints, transparency, and informed consent. It also takes into account the potential impact of DCD on all stakeholders, including innocent users, and aims to minimize harm while maximizing the benefits of cyber deception. By adopting an ethical framework, designers of DCD can ensure that their actions align with ethical principles and promote the greater good. 

The first principle of the doctrine is called the good-will principle which mandates that any defender designing DCD must have a good intention for the social good of network security. This principle requires the designer to prioritize the security of network users above all else. It is not enough to simply protect the network from potential threats; the designer must also ensure that the DCD does not engage in any activities that could harm innocent users.

To achieve this goal, the principle explicitly forbids the creation of ill-intended data collection and malicious use of innocent users. This means that the designer must not use the DCD to collect sensitive data from innocent users without their consent or knowledge. Additionally, the designer must not use the DCD to manipulate or harm users in any way, such as by sending misleading messages or leading them to believe that they are accessing legitimate resources when, in fact, they are being directed to malicious sites. By adhering to this principle, the designer can ensure that the DCD is used only for its intended purpose: to protect the network and its users from potential threats. Moreover, this principle can help build trust between the users and the designer, as users will be more likely to use and recommend a security solution that is designed with their best interests in mind. 

The second principle is the principle of deontology which states that the designer should incorporate deontology into the design framework that aligns well with security practices. This principle recognizes that security practices must be guided not only by technical considerations, but also by deontological principles. For instance, it would be considered unethical if the DCD sends misleading messages that could trick innocent users into clicking on the wrong links or accessing honeypots. While such tactics may be effective for the attacker, they must be deontologically constrained to a permissible set. This means that the designer must balance the need for security with the ethical considerations of using deception to achieve that security. To achieve this balance, the designer must ensure that the DCD adheres to a set of deontological principles. These principles may include obligations to respect user privacy, avoid causing harm, and ensure that users are fully informed about the purpose and operation of the DCD. For example, the designer may choose to use techniques such as warnings and disclosures to inform users about the presence and purpose of the DCD, rather than relying on deceptive tactics. By adhering to this principle, the designer can ensure that the DCD is designed with ethical considerations in mind, and that it operates in a manner that is consistent with these considerations. This can help build trust with users and other stakeholders, and can help ensure that the DCD is effective in achieving its security objectives while avoiding harm to innocent users.

The third principle of the doctrine is the no-harm principle, which states that the designer should create a design that does no harm to innocent and legitimate users. This principle is based on the ethical considerations of non-maleficence, which means avoiding actions that could cause harm to others. The no-harm principle is a fundamental ethical principle that has been adopted in many fields, including medicine, law, and technology.  In the context of technology, the no-harm principle is particularly important when it comes to cybersecurity and the use of data. For instance, the designer can avoid the collection of sensitive data without user consent, as collecting such data could potentially harm innocent users if it falls into the wrong hands. In addition, the designer can take steps to ensure that users are fully informed about the use and operation of the honeypots, so that they do not accidentally or unknowingly engage with the honeypots in a way that could cause harm. Adhering to the no-harm principle can help ensure that the design of the honeypots does not cause unintended harm to legitimate users. It can also help build trust with users and other stakeholders, as users are more likely to use and recommend a security solution that is designed with their best interests in mind. By prioritizing the no-harm principle, the designer can help promote the ethical and responsible use of technology, and ensure that the honeypots are used in a manner that is consistent with the values and principles of non-maleficence.

The fourth principle is the transparency principle, which emphasizes the need for the DCD policy to be transparent to all users. While users may not necessarily need to know the exact locations of honeypots within the network, they must be aware that they are using a network with honeypots. By making users aware of the presence and likelihood of honeypots, they can enter the network with consent to the DCD mechanisms, even for illegitimate users. This principle upholds the rights of users to know the situation they are in, rather than being caught off guard by unexpected consequences of the honeypot system.

Transparency is also a powerful deterrent against attackers. By announcing the deployment of honeypots, attackers are made aware of the risk of being caught. As a result, they may be less likely to enter the network or may adjust their attack strategy to avoid detection. Additionally, the knowledge of honeypots can help to discourage potential attackers from even attempting to attack the network in the first place, as the risk of being caught is already known.

There are several ways to implement the transparency principle in honeypot design. One approach is to include a clear warning message on the login or access page of the network, informing users of the presence and likelihood of honeypots. This message can explain the purpose of the honeypot system and the potential consequences of engaging in malicious activity. Another approach is to include a message in the network's terms and conditions or acceptable use policy that explicitly mentions the use of honeypots. This approach ensures that users are informed of the honeypot system before they even enter the network.

The fifth principle is the fairness principle, which emphasizes the need for the designer to consider all types of users, both legitimate and illegitimate. It is not enough for the designer to solely focus on designing a deceptive network system to detect or deter attackers, as this may lead to unintended consequences for legitimate users. The designer must also take into account the possibility that legitimate users may be impacted by the honeypot system itself or by the actions of attackers. Within the group of legitimate users, there are different types of users, including expert users and less experienced users. Expert users are well-versed in honeypot systems and can carefully navigate around them. They may even use honeypots for research or educational purposes. However, less experienced users may unknowingly access or use honeypots due to uncareful network activity or accidental clicks. These users are especially vulnerable to the potential negative consequences of honeypot systems.

To ensure that the honeypot system is fair to all users, the designer should invoke Rawls' veil of ignorance. Rawls' veil of ignorance is a design approach in which the designer does not know what type of user they will be. This approach ensures that the designer aims to create a honeypot system that is acceptable to all types of users, regardless of their expertise. One approach to achieving fairness is to implement access controls that limit the number of connection attempts or require authentication. This can help to ensure that legitimate users are not mistakenly identified as attackers and that the honeypot system is not overly burdensome on legitimate users. Additionally, the designer should consider the possibility of false positives, where legitimate users are incorrectly identified as attackers. The designer should implement mechanisms to minimize false positives while still maintaining the effectiveness of the honeypot system.

The principles of the doctrine are summarized in the following.

\noindent{\bf Principle I (Goodwill Principle)}: The defender must act with a sincere desire to protect the network and its users.

\noindent{\bf Principle II (Deontology Principle)}: The defender must make every effort to abide by as many deontological rules as possible, so that the use of deception is justified only when there is no alternative course of action available.

\noindent{\bf Principle III (No-Harm Principle)}: The defender must ensure that the use of deception does not harm any legitimate users of the network, and must take measures to prevent any unintentional harm.

\noindent{\bf Principle IV (Transparency Principle)}: The defender must ensure that the rules and mechanisms of the DCD are transparent to all users, such that they are aware of the use of deception in the network and can voluntarily consent to its use.

\noindent{\bf Principle V (Fairness Principle)}: The defender must design the DCD mechanisms while considering the perspective of all potential users, both legitimate and illegitimate, and should not create mechanisms that are unfair or harmful to any particular group of users.

We call an agent guided by the doctrine as {\it homo cyberus}. The DCE is an ethics principle that aims to guide ethical decision-making where the effects of technology are complex and multifaceted. It is often invoked in cases where a person or organization must choose between two courses of action, each of which has both positive and negative consequences that are mediated or impacted by technology. In this way, it is comparable to the  Doctrine of Double Effect (DDE) \cite{foot1967problem,mangan1949historical,quinn1989actions}, studied by many ethics philosophers, including  Thomas Aquinas, Joseph Mangan, and Philippa Foot. The DDE is a philosophical principle that is often used in ethical and moral discussions. It provides a framework for evaluating the ethical consequences of an action that has both good and bad effects, and it is often invoked in cases where a person must choose between two courses of action, each of which has both positive and negative consequences.

%BETTER ARGUMENT: HOW DDE AND DCE ARE DIFFERENT?

The DDE holds that an action can be morally permissible, even if it has negative consequences, as long as the action itself is not inherently wrong and the negative consequences are unintended and not the primary goal of the action. According to the DDE doctrine, the moral quality of an action should be based on the intention of the actor, rather than on the foreseen but unintended consequences. We argue that the DDE can be used to justify a defender's use of deception in the context of cybersecurity. The primary intention is to safeguard the network and thwart attackers, while any incidental harm caused to innocent users is unintended and not the desired outcome. The application of the DCE reinforces this position, particularly with regard to cyber-related scenarios. It is essential that no harm is done, and that all users -- especially the DCD -- are fully informed and give their consent.

%For example, DDE argues that the use of deception is ethical as long as the outcome of privacy breaches and harm is unintended. However, the harmful outcomes can be anticipated and prevented through a design that can mitigate the harm. Otherwise, there is no meaning for creating a repeated harmful consequences despite unintended. The cost of using such DCD would be huge as the number of agents grows. The direct application of DDE will not work. 

In the context of designing DCD, there are several reasons why DDE may not be directly applicable. Firstly, the designer of DCD faces an unknown set of agents and situations, including both insiders and attackers. The behaviors that can be justified under DCE may be deemed unethical under DDE. Therefore, the goal of the doctrine is not just to justify an action between idiosyncratic individuals, but rather to create a policy that can be used by a diverse population of agents with varying attributes and interests. Secondly, DDE has a common criticism that it is hard to distinguish between intended and merely foreseen consequences. In the context of designing DCD, the designer is sufficiently empowered to anticipate harmful consequences. Therefore, the ethical issues are not due to unanticipated consequences, but rather how to contain them.

Thirdly, it is essential for the designer to prevent harmful outcomes as much as possible, as repeated harmful consequences would render the design of DCD meaningless. Since DCD faces a large population of users and attackers, it is unethical to create harmful consequences even if they can be justified through DDE. The cost of such DCD would be significant and unscalable as the number of agents grows.
Fourthly, DCD may be too permissive in allowing actions that have harmful consequences. Therefore, principles such as deontology and the no-harm principle aim to constrain the set of morally permissive actions and create a policy that should not cause harm to others, even if doing so would result in some good consequences.
Finally, any mechanism or strategy employed in DCD must be equitable and unbiased, regardless of whether the users are known or unknown and their type. This is essential to ensure that the design of DCD aligns with ethical considerations and is fair to all stakeholders.

\section{Consistent Ethics Reasoning Frameworks}\label{sec:reasoning}

\begin{figure}[h]
\centering
 \includegraphics[width=.6\linewidth]{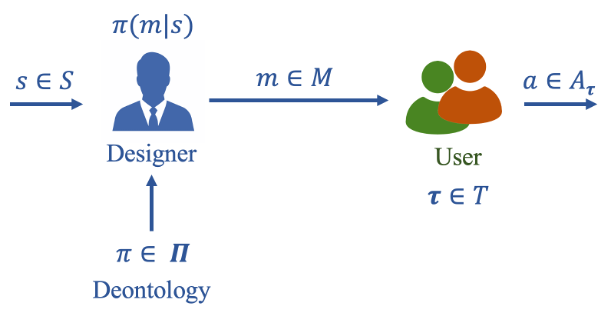}
 \caption{A game-theoretic ethical DCD reasoning model: The designer knows the true state of the world and commits to a defensive policy that is overtly announced. The goal of the policy is to induce compliance of the users while deterring the insider threats. All the insiders are aware of the policy and react to the message or recommendations from the designer. An ethical design satisfies the axioms by taking into account deontological rules and protecting the non-malicious users. }
\label{fig:ged}
\end{figure}

The DCE principle, which guides the design of cyber defense mechanisms (DCD), can be supported by consistent ethical reasoning models. In this section, we propose a stylized game-theoretic model, called Games of Ethical Duplicity (GED), which adheres to the DCD principles and provides a quantitative approach for their formal design. Figure \ref{fig:ged} illustrates the interactions between two parties: the designer or defender, and the users, including both legitimate and illegitimate (i.e., attackers) ones. The designer has access to a true state of the world, denoted by $s\in S$, which describes the cyber network state, including security posture, network traffic, and access profiles. The designer can generate signals and messages, denoted by $M$, that are broadcast to or observable by all users. In cases where the message space is the same as the state space (i.e., $M=S$), the designer can adopt a Kantian approach by conveying the true state of the world ($m=s$), or distort the message to misrepresent the true state. For instance, in the design of honeypot systems, the designer can configure honeypots to behave like production systems, creating fake network traffic flows that can be observed by the users. After observing the message, users can reason about the true state of the world and then make a decision to take action $a\in A$, where $A$ is the action set of the users. Users are assumed to have two types: legitimate ($\tau=L$) or illegitimate ($\tau=I$). The reasoning that users can do depends on their level of rationality, and the classical assumption is to view users as Bayesian agents \'a la Harsanyi \cite{harsanyi1967games} or Aumann \cite{aumann1987correlated,aumann1995epistemic}. Recent experimental justifications also support Bayesian reasoning for agents. The prior distribution of the state of the world, denoted by $p(\cdot)$ over the state space $S$, is assumed to be common knowledge and is known to all agents.

In the context of GED, the designer's objective is to minimize the risk of the network system by devising a security policy. The designer's intention can be interpreted as an act of good will, in line with Principle I. The security policy is represented as $\pi(m|s)$, which is the probability of sending message $m$ when the network is in state $s$. This policy is made public to all users in advance. The DCD is overt, meaning that the defender openly acknowledges the use of deception, and the deception policy is communicated to users before they participate in the network. This approach adheres to Principle IV. The designer's moral budget dictates the set of choices of $\pi(m|s)$, subject to deontological constraints. For instance, it would be deemed by designers at certain jurisdictions that it is unethical to mislead innocent users into accessing risky nodes by misinforming them that the network is secure. Furthermore, certain state information needs to be disclosed truthfully when it is life-critical. The set of all allowable strategies that satisfy the defender's moral budget is denoted as $\Pi$, which ensures consistency with Principle II. 

Deontological rules can apply not only to designers, but also to users. In DCD design, agents may be prohibited from certain actions in a given state or for all states, creating categorical prohibitions. An agent-relative obligation is a requirement for a particular agent to take or refrain from an action. However, such obligations can be prohibitively costly. In some cybersecurity applications, poorly designed deontological rules can compromise usability. Therefore, the issue of moral budget arises again, and it becomes important to define deontological principles that strike the right balance between security and usability.

The selection of $\Pi$ must be consistent with the principles of the doctrine. Specifically, it is crucial that $\Pi$ is chosen with the aim of minimizing the distortion between $m$ and $s$. If a designer intentionally limits their design to a subset of misinformational policies, they could violate the Goodwill or No-harm principle of the doctrine and thus compromise the ethical nature of the design. Therefore, the designer's intent should not be evaluated solely based on the objective of the design, but also on the choice of deontology. The deontological constraints can be seen as the dual of the designer's intent. The primal view of the intent is to attain maximum security within the boundaries of deontology. The dual view of the intent is to use the minimum set of deontological boundaries to achieve satisfactory security. These two views can be mathematically formalized into primal and dual problems in optimization theory.

Legitimate users are those who have authorized access to a network. In an enterprise environment, the primary objective of a DCD is to safeguard the cyberspace, providing legitimate users with a secure and risk-free environment. The use of DCD focuses on countering attackers, but it may also have unintended consequences, such as trapping certain legitimate users into harmful situations, resulting in credential revocation, personal information collection, or victimization by illegitimate users. To address this issue, the GED model categorizes legitimate users based on their attributes in a specific context. For example, in the study of insider threats, legitimate users may be classified as negligent users, compromised users, or influenced users. Negligent users are those who inadvertently take careless actions. Compromised users are victims of an attack, such as phishing or social engineering, that allows an external attacker to gain access to the network and take actions like a legitimate user. Influenced users are those who are coerced or manipulated by external actors to act against the defender.  

The design of the GED framework considers the intent and objectives of different types of users. Firstly, it formulates policies that ensure that all users are willing to use the network and that the outcomes of their actions align with their expectations. This approach minimizes the risk of non-compliant or unexpected behavior. Secondly, the designer aims to maximize the security of everyone involved. This requires taking into account the needs and behaviors of all relevant user types, as the overall security benefit of the network depends on the security of the entire population. This approach aligns with the Fairness Principle, where the designer can also be one of the users. In this case, the veil of ignorance ensures that everyone's interests are considered and addressed, regardless of their role or position in the network.

Illegitimate users may include both external attackers and malicious insiders. In the context of overt DCD, these individuals are typically aware of the deception policies put in place by the designer. Their attack can be seen as a voluntary participation in GED, and they are willing to accept the consequences of their actions. Illegitimate users can also be considered as a new type of user, in addition to the existing legitimate user types. The GED design can incorporate this type of user by creating overt DCD policies that provide no incentives for illegitimate users to participate or deviate from their inaction. This approach can induce deterrence and promote compliance among all user types.

The objective of DCD design is to enhance the security posture of both the network and its users. It is essential that the outcome of the design exceeds that of a laissez-faire approach or, at the very least, does not cause any perceptible harm or epistemic damage to the designer. Therefore, the design will be deemed unsatisfactory if the resulting performance falls below an acceptable threshold, which we refer to as the ``no-harm threshold." This approach aligns with the No-harm Principle, ensuring that the DCD design is consistent with the ethical imperative to minimize harm.

The entire framework can be conceptually summarized using the following mathematical GED optimization problem, denoted by MGED. The mathematical formalism enables a quantitative approach to ethics analysis, making a formal design of ethical deception possible. 
\begin{eqnarray}
\textrm{(MGED)} \ \  \min_{\pi\in \Pi}& \sum_{s\in S, m\in M} U_D(s, \pi, a, q) \\
\label{IC} & U_\tau(a_\tau, m,  \pi) \geq U_\tau(a'_\tau, m,  \pi), \ \ \forall a'_\tau \in A_\tau, \tau \in T \\
\label{IR} & U_\tau(a_\tau, m,  \pi) \geq \underline{U}_\tau, \ \ \forall \tau \in T
\end{eqnarray}

Here, $a:=\{a_\tau\}_{\tau\in T} \in A$, where $A:=\prod_{\tau\in T} A_\tau$, is the action profile of the users, including legitimate and illegitimate ones; $q:=\{q_{\tau}\}_{\tau\in T} \in Q$ is the population profile of the users, with $q_\tau$ denoting the probability of meeting a user of type $\tau$ in the population and $Q$ is a $|T|-$dimensional simplex. The pair $(a,q)$ is called a user profile. $U_D: S\times \Pi\times A\times P\rightarrow \mathbb{R}$ is the risk function of the network, which aims to capture the risk of network under the policy $\pi$ and the user profile $(a,q)$. $U_\tau: A_\tau \times M \times \Pi \rightarrow \mathbb{R}$ is the utility of the user of type $\tau$ when the user takes an action $a_\tau \in A_\tau$ after observing a message $m\in M$ under the overt policy $\pi\in \Pi$. Users' utility function captures their intent, the reasoning based on the observation, and the consent to the overt DCD policy $\pi$. Utility functions can take various forms depending on the type of user. Legitimate users have an intention that aligns with the designer, whereas illegitimate users may have conflicting objectives. In the second constraint (\ref{IR}), $\underline{U}_\tau$ represents the user's threshold utility level, which captures their minimum expectation. Users will only participate in the game voluntarily if they can achieve a utility level that meets or exceeds their expectations. If an illegitimate user foresees that they will be worse off by attacking the network, they will leave the network. 

The reasoning process of the users upon the observation is subsumed in the utility function. The MGED can be further instantiated for certain applications such as insider threat. For example, in \cite{huang2021duplicity}, Huang and Zhu have customized the GED framework into a variant of Bayesian persuasion problem.  They represent utility functions with matrices for discrete sets of actions, and the designer's objective is to encourage compliant behavior that promotes strong security practices. In this setup, users' reasoning processes are Bayesian, and their utility is determined by expected utility under posterior, i.e.,
$$p(s|m)= \frac{\pi(m|s)p(s)}{\sum_{m\in M}\pi(m|s)p(s)}.$$

This assumption is contextualized in the work of \cite{huang2021duplicity}, which develops a GED framework to design deception mechanisms that consist of a generator, an incentive modulator, and a trust manipulator, referred to as the GMM mechanism. It creates a mathematical programming problem to determine the optimal GMM mechanism, assess the maximum enforceable security policies, and establish conditions for user identifiability and manageability to facilitate cyber attribution and user management. This optimal GMM mechanism can encourage positive behavior from both selfish and adversarial insiders, thus enhancing the overall security of the insider network. The rationality of the reasoning  can take other forms to account for bounded rationality of the decision-making \cite{dhami2016foundations}, information structure \cite{li2022role}, and epistemology \cite{perea2012epistemic}.

While the reasoning framework introduced in this section is based on GED, there are various other reasoning frameworks that can also be used to ensure consistency with the doctrine. Game-theoretic frameworks, such as signaling games \cite{pawlick2018modeling,xu2015cyber}, Bayesian games \cite{zhao2020finite,zhang2020game}, and dynamic games \cite{horak2016point,horak2017manipulating}, have proven to be powerful tools for developing effective deception strategies in various domains, including cyber deception for communication networks \cite{clark2012deceptive,zhu2012deceptive}, cyber-physical systems \cite{zhu2018multi,rass2017physical}, and critical infrastructures \cite{chen2019game,huang2018factored,rass2020cyber}. However, these frameworks often overlook the ethical implications of their use, leading to potentially morally ambiguous and harmful situations.  By making or choosing a game-theoretic framework consistent with the DCE, we can extend their scope beyond mere utilitarianism and efficiency to include broader ethical considerations and ensure that the use of these frameworks does not result in harmful consequences or violate fundamental ethical principles.

Formal logic representations, for instance, have been proven to be effective tools for synthesizing and verifying network security policies. In \cite{govindarajulu2017automating}, deontic cognitive event calculus was used to formalize the principles of the doctrine of double effect and automate the design of autonomous systems that comply with it. Similarly, in \cite{lindner2020evaluation}, linear temporal logic was employed to formalize ethical principles and evaluate the moral permissibility of a plan. These different forms of formalism offer the opportunity to tailor reasoning frameworks to specific contexts and provide corresponding foundations for analyzing complex systems. By using a variety of formalisms, we can gain a better understanding of the context in which the reasoning framework is being applied and improve our ability to reason about complex systems. Furthermore, the extension of the reasoning framework to include different formalisms allows us to take a more holistic approach to analyzing complex systems, incorporating different types of reasoning and diverse perspectives. This can result in more effective and efficient security policies and systems that are better suited to the specific needs of different organizations and contexts.

\subsection{Existence of Ethical Solutions}

Using a reasoning framework can be a valuable tool for verifying the permissibility of a solution, even if it is not always effective for finding one. However, uncertainty may arise about the existence of a feasible solution that aligns with relevant ethical principles. To address this, consistent reasoning frameworks can be employed to identify such solutions. However, in situations where multiple ethical principles must be satisfied, the reasoning frameworks may not produce any feasible solutions. For example, strictly adhering to Kantian ethics can lead to impractical solutions, while taking a purely utilitarian perspective can result in moral dilemmas and ethical concerns.

When a consistent reasoning framework fails to yield a solution, it is not necessarily the case that there are no ethical solutions available. Instead, it may be worth exploring different reasoning frameworks to identify alternative solutions. 
It is useful to leverage the reasoning framework to identify fundamental tradeoffs that balance morality and functionality. Striking a middle ground that accommodates both can facilitate effective decision-making. When the fundamental tradeoffs obtained from a framework align with our intuition and experiences in real-world applications, it can help justify the choice of the reasoning framework.

\subsection{Inconsistent Reasoning Frameworks}
As an alternative framework, we can consider a reasoning framework that differs significantly from the standard approach by disposing of the commitment or announcement of the policy $\pi(m|s)$. In other words, users are not provided with ex ante knowledge of the DCD policy, and they can only respond to observed messages. This can leave users uncertain about what to expect when using the network, requiring significant trial and error to learn the right behaviors. During this learning process, users may make errors and mistakes, which can be costly and harmful due to the spillover effect of the DCD. The GED in this context is a reasoning framework for covert deception. The designer does not disclose the DCD policy, and users must observe and learn themselves. However, the lack of prior knowledge or information from the user's end can lead to regret over their participation and the violation of the transparency principle. Therefore, this framework is considered inconsistent with the doctrine and its outcome is deemed unethical.

%PRESENT TAO'S RESULT THAT COMPARES THE TWO.

\subsection{From Contextual Models to Praxis}

In \cite{huang2022zetar}, the authors delve into the praxis of the contextualized model, a framework that aims to address the challenges posed by insider threats within organizations. Specifically, they propose the use of a feedback data-driven system that continuously monitors and assesses the compliance of individuals and systems within the organization, in order to proactively detect and prevent potential security breaches. To implement this framework, the authors suggest creating an overlay of continuous monitoring, trust management, and access control protocols. This would enable the development of a software tool for the enterprise network that not only detects and responds to security threats but also enforces compliance policies and controls.

The DCE is  a powerful cyber moral principle that can guide the development of standards, laws, and policies in the digital realm. It provides a reasoning or quantitative framework that can help balance competing interests, such as securing cyberspace while also protecting users from unintended harm. By evaluating the ethical implications of designs, the DCE can assist lawmakers in developing laws that achieve a balance between these competing interests. Incorporating the DCE into the development of laws and policies can encourage ethical decision-making among individuals and organizations.

Another crucial implication of the DCE is accountability. By requiring individuals and organizations to consider the potential consequences of their actions and evaluate the ethical implications of their choices, lawmakers can develop laws that hold them accountable for their actions. The DCE's principled approach can also help address complex moral issues, such as end-of-life care, by considering the potential consequences of different actions. Overall, the DCE is a valuable tool for lawmakers in addressing ethical issues in the digital realm and ensuring that technology is developed and used in a responsible and ethical manner.

%a cyber moral principle that can guide the development standards, laws, and policies. It provides a reasoning or quantitative framework to balance competing interests such as securing the cyberspace and protecting users from unintended harm.  By evaluating the ethical implications of the designs, the DCE can help lawmakers develop laws that achieve a balance between these competing interests. By incorporating DCE into the development of laws and policies, lawmakers can encourage ethical decision-making among individuals and organizations. Accountability is another implication of the DCE.   By requiring individuals and organizations to consider the potential consequences of their actions and evaluating the ethical implications of actions, lawmakers can develop laws that hold individuals and organizations accountable for their actions. The DCE can help address complex moral issues by providing a principled approach to evaluating the moral implications of actions. By incorporating the DCE into lawmaking, lawmakers can address complex moral issues, such as end-of-life care, by considering the potential consequences of different actions.

\section{Broader Applications of the DCE}\label{sec:broader}
The application of DCE extends beyond the realm of cyber deception to other ethical problems in the cyberspace.  We discuss two relevant cases. Both are motivated by the disputes over recent event. The first case pertains to the application of DCE in the context of online recommendation systems used by video-sharing platforms such as YouTube. These platforms use complex algorithms to match new videos with potentially interested viewers. However, recent disputes have arisen over allegations that YouTube has promoted violent Islamist ideology by recommending videos that promote ISIS and violence. This has raised ethical concerns about the responsibility of these platforms to prevent the spread of extremist content online \cite{murthy2021evaluating,downing2021memeing}. To further explore the ethical implications of such practices, we can consider a variant of this problem. For instance, suppose that YouTube recommends a video that promotes a health product that causes chronic disease. In this scenario, several ethical questions arise. %One case of its application is in the context of online recommendation systems, such as those used by YouTube, which algorithmically match new videos with potentially interested viewers.  Recent disputes over the case alleging that YouTube helped spread violent Islamist ideology have raised the question of whether it is ethical for YouTube to recommend videos that promote ISIS and violence (see \cite{murthy2021evaluating,downing2021memeing}). We can consider a variant of this problem. Suppose that YouTube recommends a video that sells a health product that causes chronic disease. To assess the ethical implications of this practice, several questions need to be considered.

Firstly, it is important to consider whether users are informed of YouTube's recommendation policies and are aware of the possibility of video recommendations. If not, then the lack of transparency violates the doctrine of deception. Secondly, the intent of YouTube's recommendation system must be scrutinized. If YouTube's primary goal is to benefit itself rather than serve the social good of its users, then the system may be deemed unethical. Thirdly, it is necessary to determine whether YouTube's recommendations are likely to cause harm to its users based on prior knowledge. If there is a reasonable likelihood that users will be harmed, then YouTube's recommendation system is unethical.  Lastly, it is important to consider whether other users may encounter similar problems due to YouTube's recommendation system. If not, then the system's design may need to be reevaluated to ensure that it complies with ethical principles. If any of the answers to these questions are negative, then it could be considered unethical for YouTube to recommend videos that promote ISIS and violence. This would be due to a violation of the DCE, which requires transparency and a goodwill intent to serve the social good of users.

The second case highlights the importance of AI ethics and accountability. It involves a thought experiment regarding COVID testing. Imagine that you have purchased a COVID test kit and used it to collect a sample from yourself. The kit sends the data to the cloud where machine learning algorithms process the information and provide you with test results in a few hours. The entire process is automated, with no human agents involved. Unfortunately, the result turns out to be positive, prompting you to report it to your organization and cancel an important business meeting. However, after using other testing methods, you discover that you do not have COVID and the AI test kit has led to the loss of a valuable project opportunity. This scenario highlights the potential consequences of AI solutions in other cases that could be even more severe. Should the AI test kit developer be held accountable for this loss? To answer this question, we must first examine whether the AI product was ethically developed. One key aspect to consider is whether the designer has informed users of the product's false positive and false negative rates through labeling and product instructions. Secondly, we must consider whether the algorithm was designed with good intentions, taking into account the variabilities of different groups. Finally, we must evaluate whether the AI reveals the truth state of the nature or obscures or hides some information. The answers to these questions will help determine whether the practice is consistent with the DCE.

\section{Discussions and Conclusions} \label{sec:conclusion}

Cyberspace is often referred to as the ``wild west" due to its lack of established rules and regulations. This is largely due to the absence of agreed-upon ethical principles, which makes it challenging to establish accountability, regulations, and laws in cyberspace. To address this, we must examine cyberspace from fundamental philosophical principles. In this context, our work is motivated by the issue of defensive deception, and we seek to examine the ethics of using it in cyberspace. To do so, we have canvassed the doctrine of cyber effect and elaborated on five ethical principles: the goodwill principle, deontology principle, no-harm principle, transparency principle, and fairness principle. We argue that this doctrine can be instantiated into a reasoning framework called the game of ethical duplicity, which is consistent with the doctrine and can be used to guide the analysis of the design of defensive cyber deception. While the scope of our doctrine is not limited to cyber deception, it can be applied to a broader range of ethical issues in various contexts. There are many other reasoning frameworks that could be studied in the future to further enhance our understanding of ethics in cyberspace. Ultimately, by examining cyberspace from a philosophical perspective, we can establish ethical principles that enable greater accountability, regulation, and protection in the digital domain.

The reflective equilibrium method, when used in conjunction with a consistent reasoning framework, offers a promising approach for developing a coherent set of principles for designing DCD in fast-changing adversarial environments, where new attack behaviors emerge and new behaviors are observed on a daily basis. Through reflection and adjustment, this approach enables the creation of a mutually consistent and belief-coherent set of principles that can be agilely adapted to emergency situations, especially in warfare contexts.

One example of the application of this approach can be found in the work of Huang and Zhu \cite{huang2022zetar}, where an iterative approach is used to update the policy within the reasoning framework based on observations of user behaviors. The method of reflective equilibrium could be integrated into this iterative policy, allowing for updates not just to the policy, but also to the principles that guide the reasoning framework underlying the policy. By using the reflective equilibrium method in this way, designers of DCD can continually refine their ethical principles, taking into account new information and changing circumstances. This approach offers a way to create a set of principles that are both coherent and responsive to emerging threats and challenges in the cyberspace environment.

\singlespacing
\setlength\bibsep{0pt}
\bibliographystyle{abbrv}
\bibliography{literature-ethics}

\begin{thebibliography}{10}

\bibitem{aumann1995epistemic}
R.~Aumann and A.~Brandenburger.
\newblock Epistemic conditions for nash equilibrium.
\newblock {\em Econometrica: Journal of the Econometric Society}, pages
  1161--1180, 1995.

\bibitem{aumann1987correlated}
R.~J. Aumann.
\newblock Correlated equilibrium as an expression of bayesian rationality.
\newblock {\em Econometrica: Journal of the Econometric Society}, pages 1--18,
  1987.

\bibitem{castiglione1995social}
D.~Castiglione, J.~Charvet, D.~Coole, and M.~Forsyth.
\newblock {\em The social contract from Hobbes to Rawls}.
\newblock Routledge, 1995.

\bibitem{cerf2011first}
V.~G. Cerf.
\newblock First, do no harm.
\newblock {\em Philosophy \& Technology}, 24(4):463--465, 2011.

\bibitem{chen2017security}
J.~Chen and Q.~Zhu.
\newblock Security as a service for cloud-enabled internet of controlled things
  under advanced persistent threats: a contract design approach.
\newblock {\em IEEE Transactions on Information Forensics and Security},
  12(11):2736--2750, 2017.

\bibitem{chen2019game}
J.~Chen and Q.~Zhu.
\newblock {\em A Game-and Decision-Theoretic Approach to Resilient
  Interdependent Network Analysis and Design}.
\newblock Springer, 2019.

\bibitem{clark2012deceptive}
A.~Clark, Q.~Zhu, R.~Poovendran, and T.~Ba{\c{s}}ar.
\newblock Deceptive routing in relay networks.
\newblock In {\em Decision and Game Theory for Security: Third International
  Conference, GameSec 2012, Budapest, Hungary, November 5-6, 2012. Proceedings
  3}, pages 171--185. Springer, 2012.

\bibitem{de2017utilitarianism}
K.~de~Lazari-Radek and P.~Singer.
\newblock {\em Utilitarianism: A very short introduction}.
\newblock Oxford University Press, 2017.

\bibitem{dhami2016foundations}
S.~Dhami.
\newblock {\em The foundations of behavioral economic analysis}.
\newblock Oxford University Press, 2016.

\bibitem{downing2021memeing}
J.~Downing.
\newblock Memeing and speaking vernacular security on social media: Youtube and
  twitter resistance to an isis islamist terror threat to marseille, france.
\newblock {\em Journal of Global Security Studies}, 6(2):ogz081, 2021.

\bibitem{foot1967problem}
P.~Foot.
\newblock The problem of abortion and the doctrine of the double effect.
\newblock 1967.

\bibitem{govindarajulu2017automating}
N.~S. Govindarajulu and S.~Bringsjord.
\newblock On automating the doctrine of double effect.
\newblock {\em arXiv preprint arXiv:1703.08922}, 2017.

\bibitem{gupta2017fighting}
B.~B. Gupta, A.~Tewari, A.~K. Jain, and D.~P. Agrawal.
\newblock Fighting against phishing attacks: state of the art and future
  challenges.
\newblock {\em Neural Computing and Applications}, 28:3629--3654, 2017.

\bibitem{harsanyi1967games}
J.~C. Harsanyi.
\newblock Games with incomplete information played by {Bayesian players, I--III
  Part I} the basic model.
\newblock {\em Management science}, 14(3):159--182, 1967.

\bibitem{horak2016point}
K.~Hor{\'a}k and B.~Bo{\v{s}}ansk{\`y}.
\newblock A point-based approximate algorithm for one-sided partially
  observable pursuit-evasion games.
\newblock In {\em Decision and Game Theory for Security: 7th International
  Conference, GameSec 2016, New York, NY, USA, November 2-4, 2016, Proceedings
  7}, pages 435--454. Springer, 2016.

\bibitem{horak2017manipulating}
K.~Hor{\'a}k, Q.~Zhu, and B.~Bo{\v{s}}ansk{\`y}.
\newblock Manipulating adversary's belief: A dynamic game approach to deception
  by design for proactive network security.
\newblock In {\em International Conference on Decision and Game Theory for
  Security}, pages 273--294. Springer, 2017.

\bibitem{huang2018factored}
L.~Huang, J.~Chen, and Q.~Zhu.
\newblock Factored markov game theory for secure interdependent infrastructure
  networks.
\newblock {\em Game Theory for Security and Risk Management: From Theory to
  Practice}, pages 99--126, 2018.

\bibitem{huang2019adaptive}
L.~Huang and Q.~Zhu.
\newblock Adaptive honeypot engagement through reinforcement learning of
  semi-markov decision processes.
\newblock In {\em International Conference on Decision and Game Theory for
  Security}, pages 196--216. Springer, 2019.

\bibitem{huang2019dynamic}
L.~Huang and Q.~Zhu.
\newblock Dynamic bayesian games for adversarial and defensive cyber deception.
\newblock In {\em Autonomous cyber deception}, pages 75--97. Springer, 2019.

\bibitem{huang2021duplicity}
L.~Huang and Q.~Zhu.
\newblock Duplicity games for deception design with an application to insider
  threat mitigation.
\newblock {\em IEEE Transactions on Information Forensics and Security},
  16:4843--4856, 2021.

\bibitem{huang2022zetar}
L.~Huang and Q.~Zhu.
\newblock Zetar: Modeling and computational design of strategic and adaptive
  compliance policies.
\newblock {\em arXiv preprint arXiv:2204.02294}, 2022.

\bibitem{huang2021reinforcement}
Y.~Huang, L.~Huang, and Q.~Zhu.
\newblock Reinforcement learning for feedback-enabled cyber resilience.
\newblock {\em arXiv preprint arXiv:2107.00783}, 2021.

\bibitem{jajodia2011moving}
S.~Jajodia, A.~K. Ghosh, V.~Swarup, C.~Wang, and X.~S. Wang.
\newblock {\em Moving target defense: creating asymmetric uncertainty for cyber
  threats}, volume~54.
\newblock Springer Science \& Business Media, 2011.

\bibitem{kant2001lectures}
I.~Kant.
\newblock {\em Lectures on ethics}, volume~2.
\newblock Cambridge University Press, 2001.

\bibitem{li2022role}
T.~Li, Y.~Zhao, and Q.~Zhu.
\newblock The role of information structures in game-theoretic multi-agent
  learning.
\newblock {\em Annual Reviews in Control}, 2022.

\bibitem{lindner2020evaluation}
F.~Lindner, R.~Mattm{\"u}ller, and B.~Nebel.
\newblock Evaluation of the moral permissibility of action plans.
\newblock {\em Artificial Intelligence}, 287:103350, 2020.

\bibitem{mangan1949historical}
J.~T. Mangan.
\newblock An historical analysis of the principle of double effect.
\newblock {\em Theological Studies}, 10(1):41--61, 1949.

\bibitem{mayer2016relevance}
B.~Mayer.
\newblock The relevance of the no-harm principle to climate change law and
  politics.
\newblock {\em Asia Pacific Journal of Environmental Law}, 19(1):79--104, 2016.

\bibitem{murthy2021evaluating}
D.~Murthy.
\newblock Evaluating platform accountability: terrorist content on youtube.
\newblock {\em American behavioral scientist}, 65(6):800--824, 2021.

\bibitem{pawlick2018modeling}
J.~Pawlick, E.~Colbert, and Q.~Zhu.
\newblock Modeling and analysis of leaky deception using signaling games with
  evidence.
\newblock {\em IEEE Transactions on Information Forensics and Security},
  14(7):1871--1886, 2018.

\bibitem{pawlick2019game}
J.~Pawlick, E.~Colbert, and Q.~Zhu.
\newblock A game-theoretic taxonomy and survey of defensive deception for
  cybersecurity and privacy.
\newblock {\em ACM Computing Surveys (CSUR)}, 52(4):82, 2019.

\bibitem{perea2012epistemic}
A.~Perea.
\newblock {\em Epistemic game theory: reasoning and choice}.
\newblock Cambridge University Press, 2012.

\bibitem{quinn1989actions}
W.~S. Quinn.
\newblock Actions, intentions, and consequences: The doctrine of double effect.
\newblock {\em Philosophy \& Public Affairs}, pages 334--351, 1989.

\bibitem{rass2017physical}
S.~Rass, A.~Alshawish, M.~A. Abid, S.~Schauer, Q.~Zhu, and H.~De~Meer.
\newblock Physical intrusion games--optimizing surveillance by simulation and
  game theory.
\newblock {\em IEEE Access}, 5:8394--8407, 2017.

\bibitem{rass2020cyber}
S.~Rass, S.~Schauer, S.~K{\"o}nig, and Q.~Zhu.
\newblock {\em Cyber-Security in Critical Infrastructures}, volume 297.
\newblock Springer, 2020.

\bibitem{rass2016gadapt}
S.~Rass and Q.~Zhu.
\newblock Gadapt: a sequential game-theoretic framework for designing
  defense-in-depth strategies against advanced persistent threats.
\newblock In {\em International conference on decision and game theory for
  security}, pages 314--326. Springer, 2016.

\bibitem{salahdine2019social}
F.~Salahdine and N.~Kaabouch.
\newblock Social engineering attacks: A survey.
\newblock {\em Future Internet}, 11(4):89, 2019.

\bibitem{sharpe1997no}
V.~A. Sharpe.
\newblock Why do no harm?
\newblock {\em Theoretical medicine}, 18:197--215, 1997.

\bibitem{spitzner2003honeypots}
L.~Spitzner.
\newblock {\em Honeypots: tracking hackers}, volume~1.
\newblock Addison-Wesley Reading, 2003.

\bibitem{van2019cognitive}
A.~Van Der~Heijden and L.~Allodi.
\newblock Cognitive triaging of phishing attacks.
\newblock In {\em USENIX Security Symposium}, pages 1309--1326, 2019.

\bibitem{vermeule2001veil}
A.~Vermeule.
\newblock Veil of ignorance rules in constitutional law.
\newblock {\em Yale LJ}, 111:399, 2001.

\bibitem{winfield2019ethical}
A.~Winfield.
\newblock Ethical standards in robotics and ai.
\newblock {\em Nature Electronics}, 2(2):46--48, 2019.

\bibitem{xu2015cyber}
Z.~Xu and Q.~Zhu.
\newblock A cyber-physical game framework for secure and resilient multi-agent
  autonomous systems.
\newblock In {\em 2015 54th IEEE Conference on Decision and Control (CDC)},
  pages 5156--5161. IEEE, 2015.

\bibitem{zhang2020game}
T.~Zhang, L.~Huang, J.~Pawlick, and Q.~Zhu.
\newblock Game-theoretic analysis of cyber deception: Evidence-based strategies
  and dynamic risk mitigation.
\newblock {\em Modeling and Design of Secure Internet of Things}, pages 27--58,
  2020.

\bibitem{zhao2020finite}
Y.~Zhao, L.~Huang, C.~Smidts, and Q.~Zhu.
\newblock Finite-horizon semi-markov game for time-sensitive attack response
  and probabilistic risk assessment in nuclear power plants.
\newblock {\em Reliability Engineering \& System Safety}, 201:106878, 2020.

\bibitem{zhu2015game}
Q.~Zhu and T.~Ba\c{s}ar.
\newblock Game-theoretic methods for robustness, security, and resilience of
  cyberphysical control systems: games-in-games principle for optimal
  cross-layer resilient control systems.
\newblock {\em Control Systems, IEEE}, 35(1):46--65, 2015.

\bibitem{zhu2013game}
Q.~Zhu and T.~Ba{\c{s}}ar.
\newblock Game-theoretic approach to feedback-driven multi-stage moving target
  defense.
\newblock In {\em International conference on decision and game theory for
  security}, pages 246--263. Springer, 2013.

\bibitem{zhu2012deceptive}
Q.~Zhu, A.~Clark, R.~Poovendran, and T.~Ba{\c{s}}ar.
\newblock Deceptive routing games.
\newblock In {\em 2012 IEEE 51st IEEE Conference on Decision and Control
  (CDC)}, pages 2704--2711. IEEE, 2012.

\bibitem{zhu2018multi}
Q.~Zhu and S.~Rass.
\newblock On multi-phase and multi-stage game-theoretic modeling of advanced
  persistent threats.
\newblock {\em IEEE Access}, 6:13958--13971, 2018.

\end{thebibliography}

%
%\clearpage
%
%\onehalfspacing
%
%\section*{Tables} \label{sec:tab}
%\addcontentsline{toc}{section}{Tables}
%
%
%
%\clearpage
%
%\section*{Figures} \label{sec:fig}
%\addcontentsline{toc}{section}{Figures}

%\begin{figure}[hp]
% \centering
% \includegraphics[width=.6\textwidth]{../fig/placeholder.pdf}
% \caption{Placeholder}
% \label{fig:placeholder}
%\end{figure}

%\clearpage
%
%\section*{Appendix A. Placeholder} \label{sec:appendixa}
%\addcontentsline{toc}{section}{Appendix A}

\end{document}